\documentclass[utf8]{frontiersFPHY} 

\setcitestyle{square} 
\usepackage{url,hyperref,lineno,microtype,subcaption}
\usepackage[onehalfspacing]{setspace}
\usepackage{xcolor}
\usepackage{gensymb}

\def\keyFont{\fontsize{8}{11}\helveticabold }
\def\firstAuthorLast{H. Monteiro {et~al.}} 

\def\Authors{ H. Monteiro\,$^{1}$, D. A. Barros\,$^{2}$, W. S. Dias\,$^{1}$ and J. R. D. L\'epine\,$^{3}$}

\def\Address{
$^{1}$Instituto de F\'isica e Qu\'imica, Universidade Federal de Itajub\'a, Av. BPS 1303 Pinheirinho, 37500-903 Itajub\'a, MG, Brazil\\

$^{2}$Recife, Brazil\\

$^{3}$Universidade de S\~ao Paulo, Instituto de Astronomia, Geof\'isica e Ci\^encias Atmosf\'ericas, S\~ao Paulo,  SP, Brazil}

\def\corrAuthor{H. Monteiro}
\def\corrEmail{hmonteiro@unifei.edu.br}

\begin{document}
\onecolumn

\title[The distribution of open clusters]{The distribution of open clusters in the Galaxy}

\author[\firstAuthorLast ]{\Authors} 
\address{} 
\correspondance{} 

\extraAuth{}

\maketitle

\begin{abstract}
In this work we explore the new catalog of galactic open clusters that became available recently, containing 1750 clusters that have been re-analysed using the Gaia DR2 catalog to determine the stellar memberships.
We used the young open clusters as tracers of spiral arms  and determined the spiral pattern rotation speed of the Galaxy and the corotation radius, the strongest Galactic resonance. The sample of open clusters used here increases the last one from Dias et al. (2019) used in the previous determination of the pattern speed  by dozens objects. In addition, the distances and ages values are better determined,  using improvements to isochrone fitting and including an updated extinction polynomial for the Gaia DR2 photometric band-passes, and the Galactic abundance gradient as a prior for metallicity.
In addition to the better age determinations, the catalog contains better positions in the Galactic plane and better proper motions. This allow us to discuss not only the present space distribution of the clusters, but also the space distribution of the clusters's birthplaces, obtained by integration of the orbits for a time equal to their age. The value of the rotation velocity of the arms ($28.5 \pm 1.0$ km\,s$^{-1}$\,kpc$^{-1}$) implies that the corotation radius ($R_c$) is close to the solar Galactic orbit ($R_c/R_0 = 1.01\pm0.08$), which is supported by other observational evidence discussed in this text. A simulation is presented, illustrating the motion of the clusters in the reference frame of corotation. 
We also present general statistics of the catalog of clusters, like spatial 
distribution, distribution relative to height from the Galactic plane, and
distribution of ages and metallicity. An important feature of the space distribution, the corotation gap in the gas distribution and its consequences for the young clusters, is discussed.
\tiny
 \keyFont{ \section{Keywords:} Open Clusters, Galaxy kinematics and dynamics, Galaxy structure, spiral arms, spiral pattern rotation speed}
 
\end{abstract}

\section{Introduction}

The open clusters are known to constitute one of the best classes of objects to investigate the Galactic structure and Stellar Dynamics, since they have relatively precise ages estimated from  isochrone fits. Recently, the \textit{Gaia} DR2 catalog (\cite{GaiaCollab2018}) released data on  more than 1 billion stars with magnitudes $G$ $\leq$ 21, with high precision astrometric and photometric data, which allows to improve the stellar membership determination and characterization of thousands of open clusters (\cite{CantatGaudin2020,CastroGinard2020,CastroGinard2019,LiuPang2019,Sim2019,Ferreira2020,Monteiro2020}),
resulting in more reliable determination of their fundamental parameters.
The data obtained on velocities, distances and ages allow us to present the analyses of the distribution of the open clusters in the solar neighborhood, exploring the birthplaces and actual positions on the Galactic plane.
Additionally, considering the clusters positions above or below the Galactic plane, we determine the solar displacement in the perpendicular direction from that plane. The present day position of the young open clusters shown in Figure \ref{fig:arms-age} reveals the spiral structure of our Galaxy in the extended solar neighborhood. 

The knowledge of the present day positions and the birthplaces of the open clusters, from the Galactic orbits integrated backward in time, allow us to revisit the value of spiral pattern rotation speed of the Galaxy ($\Omega_p$) and the Galactic corotation radius ($R_c$). The results show that the corotation radius 
is very close to the solar radius and, therefore, the solar orbit is under strong influence of this major Galactic resonance.
Although we try to avoid complicated models in this paper, we must say that the guidelines followed by our group of research to interpreting the kinematic phenomena that we observe are based on the following ideas: the spiral arms are grooves in the gravitational potential of the disk, and the grooves are due to the excess of star density inside the arms. The star density is explained by the stellar orbits that are not perfectly circular, and come close to each other in some regions of the Galactic disk. The mathematical description of the grooves and of the orbits can be found in \cite{Junqueira2013}.




In the following sections, we first describe the sample of open clusters, how it was obtained
and its main characteristics, compared to other samples, and the quality of the data (Section 2). 
In Section 3, we present the statistics of the sample and the distribution of the clusters in the Galactic plane, as well as the age and the Galactic radial distributions.
Section 4 shows the properties related to the height distribution from the plane of the Galaxy.
In Section 5, we present the method used to determine $\Omega_p$ and $R_c$.
In Section 6, we discuss the general results and we give and explanation for the nature of the 9\,kpc gap in the radial distribution of the open clusters, which is presented in Section 3. 

\section{The sample of open clusters}

The present version of the catalog contains 1750 clusters, which is the first update of the New Catalogue of Optically Visible Open Clusters and Candidates (DAML) (\cite{Dias2002}) in the \textit{Gaia} era (see all details in \citet{Dias2021}). The distribution of clusters in the Galactic disk is shown in Figure 1, with color proportional to the age, clearly indicating the arms in the solar neighborhood. 

\begin{figure}
\centering
\includegraphics[width=\textwidth]{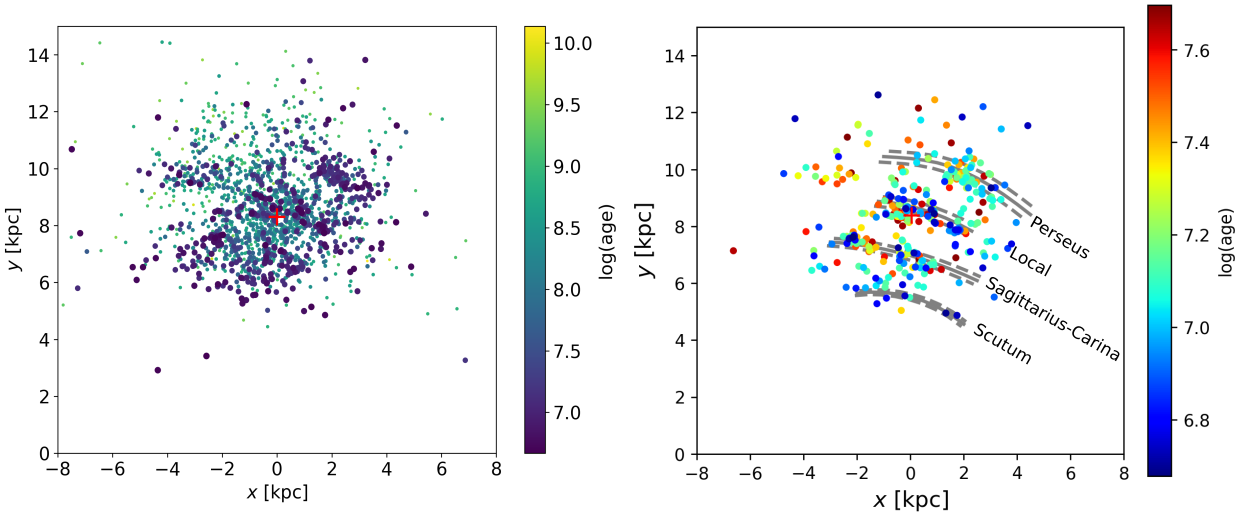}
\caption{Distribution of the 1750 open clusters in the Galactic plane. In the right plot are presented the open clusters with ages lower than 50 Myr. In light-grey are the present zero-age arms positions. The Sun (red cross) is at coordinates (0, 8.3)\,kpc and the Galactic center is at (0,0). The vector angular velocity is perpendicular to the x–y plane pointing in the direction of the paper.}
\label{fig:arms-age}
\end{figure}

In the the right plot of Figure \ref{fig:arms-age}, are presented 566 young clusters with ages lower than 50\,Myr also with colors proportional to the age. Interestingly, the
clusters associated with the Perseus arm and older than 10 Myr  are situated to
the left of the zero age arm position. In contrast, most of the clusters older than 10 Myr associated with the Sagittarius-Carina arm, and particularly with the Scutum arm, are situated to the right of the corresponding arms positions. 
Compared to the other arms, the clusters of the Local arm are more in line with the arm position. This means that we can qualitatively expect to find the corotation radius between the Perseus and Sagittarius-Carina arms, that is, close to the Sun location. In the inertial frame of reference, the clusters are rotating around the Galactic center with about the velocity given by the rotation curve. In the frame of reference of the spiral arms, the velocity of the clusters have opposite signs on the two sides of corotation. 

In this work, we selected the clusters with published individual stellar membership, determined from \textit{Gaia} DR2 astrometric data by \cite{CantatGaudin2020,CastroGinard2020,CastroGinard2019,LiuPang2019,Sim2019,Ferreira2020,Monteiro2020}, to estimate distance and ages from isochrone fittings. 
Most open clusters had stellar membership probabilities determined by \cite{cantat2018} using the UPMASK procedure \cite{Krone-Martins2014A&A...561A..57K} applied to the astrometric parameters (proper motion and parallax) considering their uncertainty and the correlations between those three parameters.
The other fraction of open clusters had their stellar membership probabilities recalculated by our group, also considering the astrometric parameters and their uncertainty and the correlations  through a variation of the classic maximum likelihood approach described in \cite{Dias2014} and \cite{Monteiro2020}.

Based on the member stars with membership probability greater than 0.50, we used \textit{Gaia} DR2 data to determine mean proper motion and mean radial velocity as well as distance and age of the open clusters. Our experience with memberships is that being more strict in the membership probability (let us say 0.70) decreases the number of stars, eliminating also a number of stars that are members, and does not result in better fits.
We adopted as the cluster proper motion the simple mean of the proper motion distribution of the member stars, and considered the standard deviation (1$\sigma$) to represent the error.

We found hundreds member stars with radial velocity in the \textit{Gaia} DR2 catalog and estimated mean radial velocities for hundreds of open clusters. To accommodate different numbers of measurements and also measurements with different errors, the mean radial velocity of each open cluster was obtained by weighting the number of measurements stars and the mean error of a single measurement, according to \cite{Barford1985}. In the final catalog, we inserted mean radial velocities for 36 clusters from APOGEE (\cite{Carrera2019}), 129 from \cite{Soubiran2018}, and 145 from DAML (\cite{Dias2002}). 
This update allowed us to obtain a sample of 327 open clusters younger than 50\,Myr with a complete set of data, including mean proper motion, mean radial velocity, distance and age.


Accurate distances and ages are crucial for untangling the features (e.g. main sequence, turn-off, and giant branch) of the cluster in the color-magnitude diagram which clearly was improved with the astrometric membership determined from Gaia DR2 data. 
In this work, to determine the distances and ages of the clusters, we applied the isochrone fittings of open cluster photometric data with a global optimization algorithm, which generates a set of possible isochrone solutions (simulated open clusters) given a pre-defined initial mass function, binary fraction and metallicity. Each generated solution is then compared to the observed data through an objective function. It selects $10\%$ of the best solutions to generate a new set of solutions and the process iterates until convergence. The procedure uses the stellar membership probability to guide the isochrone fit performed by our global optimization tool which avoids the need of performing fits visually and, thus, removes most of the related subjectivity, and that allows us to obtain error estimates of the fundamental parameters via Monte-Carlo technique. For a detailed description of the cross-entropy optimization technique used, see \cite{Monteiro2020}. In this task, we used {\it Gaia} DR2 $G_{BP}$ and $G_{RP}$ magnitudes and the updated extinction polynomial for the {\it Gaia} DR2 photometric band-passes and the Galactic abundance gradient as a prior for metallicity (see details in \cite{Monteiro2020}).  

We do not expect  significant changes in the parameters distances and ages in our sample with the  {\it Gaia} DR3 release, since the {\it Gaia} distances were used to identify membership, and very few changes are expected in membership. The final distances in the catalog are derived from photometry. However, we must wait for the DR3 data
analyses to have a meaningful comparison.

\begin{figure}
\centering
\includegraphics[width=\textwidth]{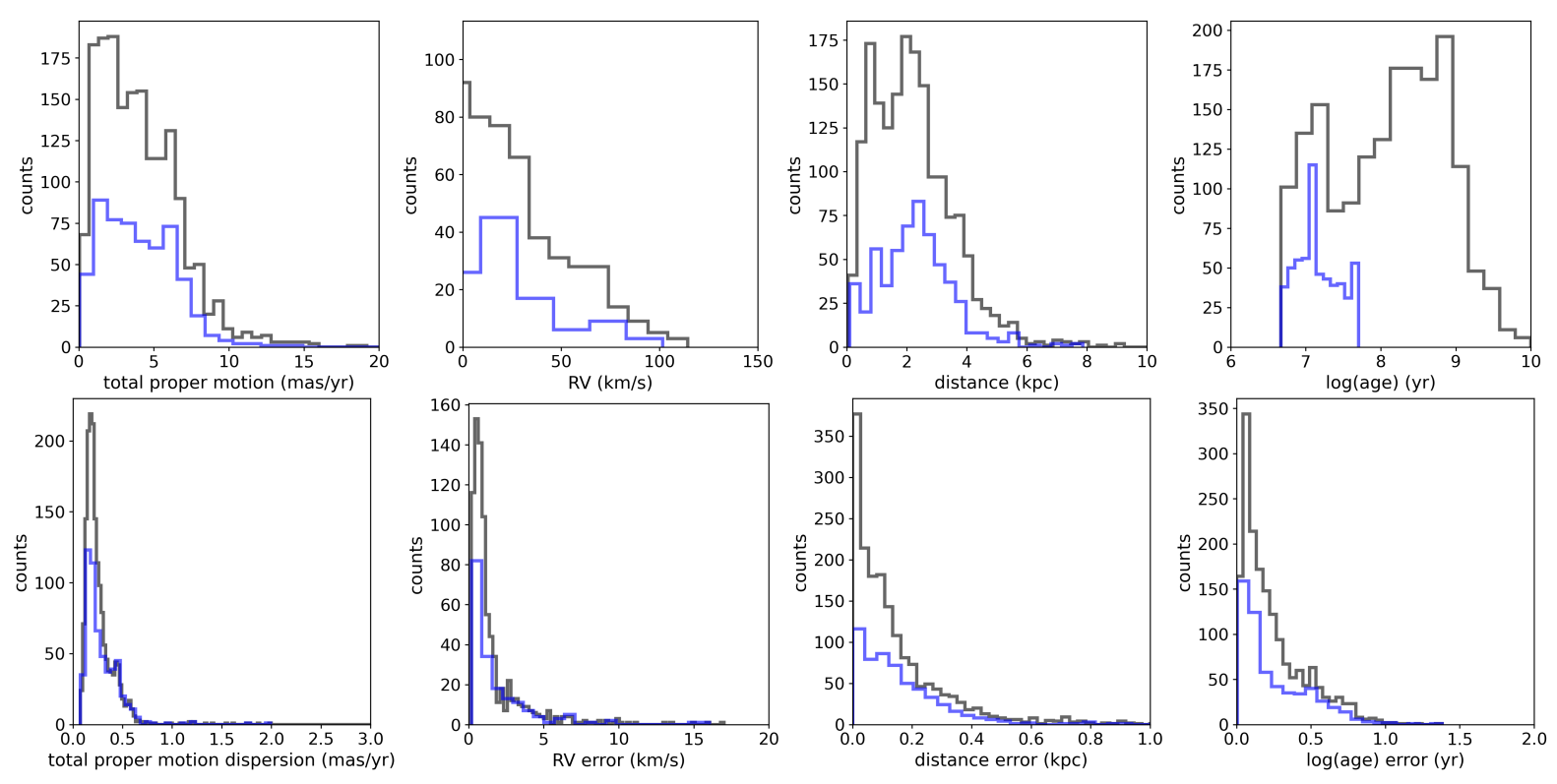}
\caption{Distribution of the total proper motion, radial velocity, distance and age from the isochrone fits and the respective uncertainties of the open clusters used in this study. The black line refers to the whole sample and the blue line refers to open clusters with age less than 50 Myr.}
\label{fig:parameters-distribution}
\end{figure}

Figure \ref{fig:parameters-distribution} shows the distribution of the total proper motion ($\sqrt{(\mu_{\alpha} \cos \delta)^2 + (\mu_{\delta})^2}$), radial velocity, distance and age of the whole sample (in black) and of the sample of young clusters (in blue) used to determine $\Omega_p$. In the lower panels we show the distribution of internal uncertainties of the same quantities. \\ 

\section{Distribution of the clusters in the Galactic plane}

The 2-D  distribution of the observed open clusters of our sample in the Galactic plane is shown in Figure \ref{fig:arms-age}. The known clusters are situated in a small portion of the Galactic plane within about 5\,kpc from the Sun. 
In the left panel of the figure, it can be noted that the distribution is more extended towards the Galactic anticenter direction (about 5\,kpc) than in the direction of the center (about 3\,kpc). This can be attributed to the dust extinction, which is stronger in the first and fourth quadrants\footnote{\textcolor{red}{quadrant I: $0\degree<l<90\degree$;  quadrant II: $90\degree<l<180\degree$; quadrant III: $180\degree<l<270\degree$ and quadrant IV: $270\degree<l<360\degree$, where l is the Galactic longitude.}} of the Galactic disk (eg. \cite{Amores2005}).\\

In the right panel of Figure \ref{fig:arms-age}, we present the open clusters with ages lower than 50\,Myr. This sample of young objects clearly follows the spiral pattern in the Solar neighborhood, according to the spiral arms log-periodic functions fitted by \cite{Reid2014}.

\begin{figure}[h]
\centering
\includegraphics[width=\textwidth]{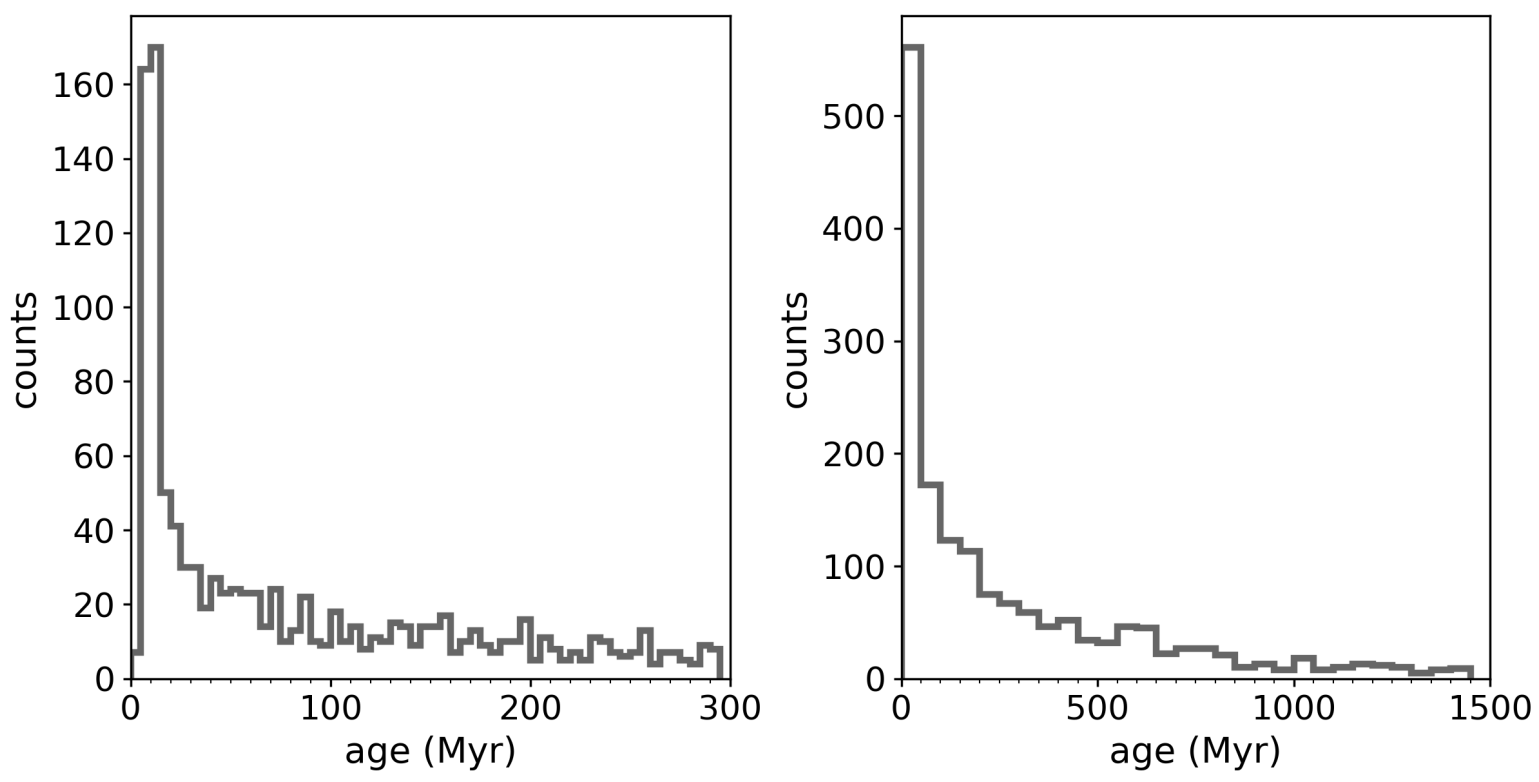}
\caption{The age distribution of the open clusters of our total sample. Two time-scales are presented, with bin width 5 Myr, extending to 300 Myr (left plot) and with bin width 40 Myr, extending to 1500 Myr (right plot). The evolution of the number of clusters is quite smooth on the two scales.} 
\label{age-distribution}
\end{figure}

In Figure \ref{age-distribution}, we present the age distribution of the clusters, in a linear scale of ages. There are a large number of clusters with age smaller than 15\,Myr, followed by a sharp decrease of the cluster number density with age.  This means that a large number of clusters do not survive longer than 10-15 Myr after their birth. Let us call this phenomenon a "high mortality" of the clusters; this could be due to: high collision rates inside the arms, the gravitational  potential gradients at the edge of the arms, or to the relaxation process, described by \cite{Binney}, (chapter 8), due to clusters that were not born with the more massive stars at the center.
The death rate later decreases; from about 30\,Myr, when most clusters leave the spiral arm in which they were born, to the end of the age interval that we investigated, there is only a  very slow death rate. The clusters are no longer connected with the spiral arms where they were born, and are travelling quietly in the inter-arm regions. The slow death rate is possibly due to evaporation, a process described by \cite{Binney}, (chapter 8). In addition to this, there can be member stars kicked off from the clusters by some gravitational interactions with objects of the Galactic disk. Apparently, no dramatic event, like a collision with a dwarf galaxy, able to produce an increase of mortality, or on the contrary, a burst of star formation, took place within the time interval that we investigated.

\begin{figure}
\centering

\includegraphics[width=\textwidth]{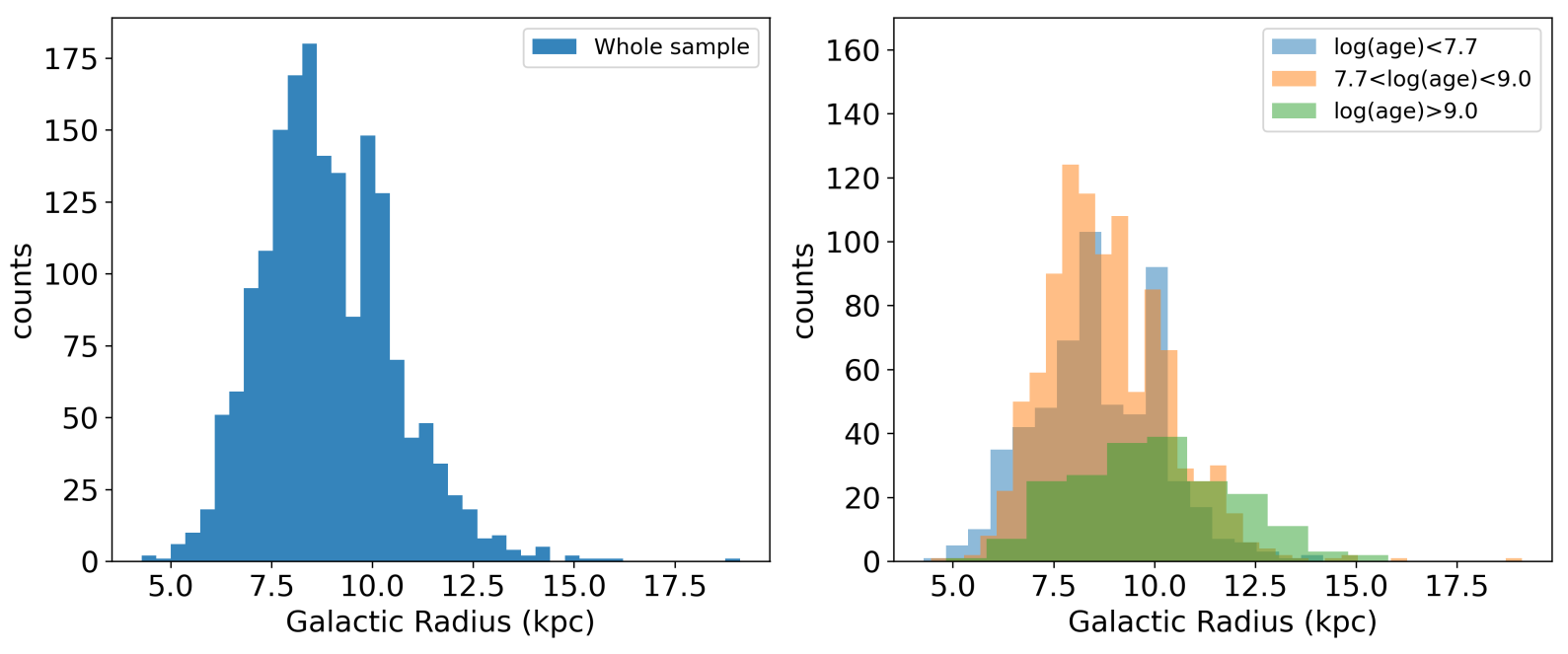}

\caption{The Galactocentric radial distribution of the open clusters sample with different age intervals. In the left-hand panel, the whole sample is presented. The minimum near 9.5\,kpc is the most evident structure. In the right-hand panel, we show the radial distribution of the objects separated in three age ranges: younger than $\log(age)<7.7$, $7.7\leq \log(age) \leq 9.0$ and $\log(age)>9.0$. We can see that the gap at 9.5\,kpc is much wider (about 1\,kpc width) and deeper in the younger cluster distribution. The gap is smaller in the intermediate age distribution and disappears in the older distribution. We interpret this as being due to open cluster's radial migrations, that tends to fill the gaps with time. Note that the brown color is the effect of superposing blue and yellow.}
\label{fig:RGC-distribution}
\end{figure}

In Figure \ref{fig:RGC-distribution}, we present the distribution of
radial distances from the Galactic Center for the clusters of our sample. There is a broad peak at about 8\,kpc, the radius of the Solar orbit, which can be explained by the observational limiting distances of sources around the Sun, as already discussed to explain Figure 1. The distribution is not perfectly symmetric, being more extended in the direction of large radii. Two factors are able to  contribute to this asymmetry. One factor is that the interstellar extinction is smaller in the direction of the external quadrants (II and III) of the Galaxy (eg. \cite{Amores2013}), and the other is the presence of the Local arm, very close to the Sun, also in those external quadrants.


\section{ Distribution  perpendicular to the Galactic plane}

\begin{figure}
\centering
\includegraphics[width=\textwidth]{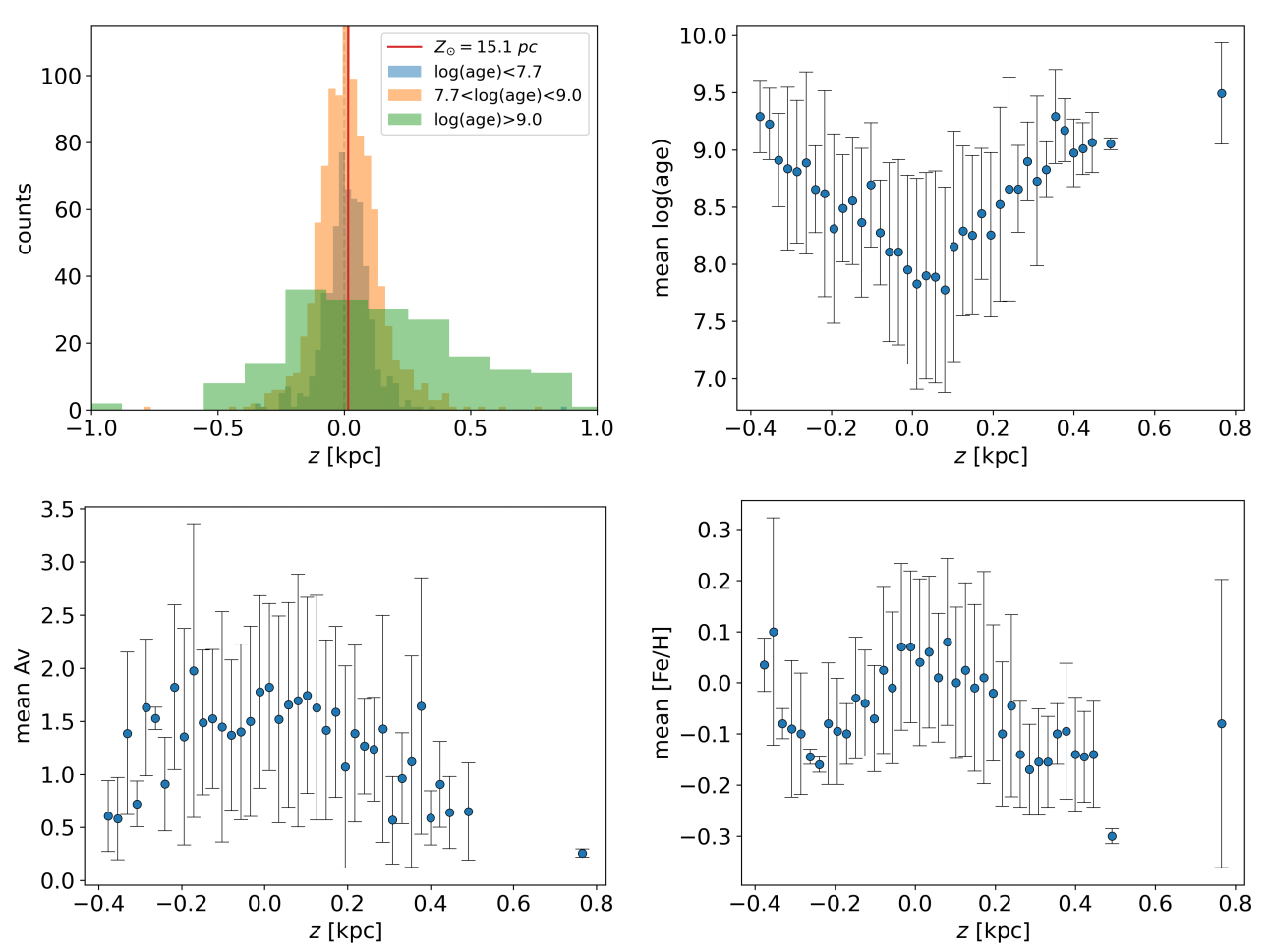}

\caption{Distribution of the open clusters and their parameters as a function of height relative to the plane of the Galaxy.The upper left panel shows histograms of the distances of the clusters to the Galactic plane in different colors  for 3  ranges of ages. Note that the brown color is the result of superposing blue and orange colors. The other panels are clearly showing the expected trends in age, extinction, and metallicity, as a function of distance to the Galactic plane.}
\label{fig:Z-distribution}
\end{figure}

The large homogeneous sample of clusters allows us to investigate and confirm many interesting properties of the structure of our Galaxy. One of these interesting parameters is the distribution of clusters as a function of height from the plane of the Galaxy. To investigate this we show, in Figure \ref{fig:Z-distribution}, the histograms of these distributions separated in three age groups: $\log(age) < 7.7$, $7.7<\log(age)<9.0$ and $\log(age) > 9.0$. Looking at these age groups we have the following statistics for the disc offset obtained from the median of the distributions and their respective standard deviation: 1) $\log(age) < 7.7$:  $\bar{z}=13.7\,\mathrm{pc};~ \sigma_z=96.6$\,pc; 2) $7.7<\log(age)<9.0$: $\bar{z}=11.9\,\mathrm{pc};~ \sigma_z=141.2$\,pc; and 3) $\log(age) > 9.0$: $\bar{z}=83.1\,\mathrm{pc};~ \sigma_z=449.4$\,pc. Looking at the sample as a whole we obtain $15.1 \pm 4.6$ \,pc for the distance of the Sun to the plane of the Galaxy. If we take only the younger clusters with $\log(age) < 7.7$ we get $15.5 \pm 4.1$ \,pc. \\ 

The values obtained of median z are in general agreement with the values obtained by other authors also using samples of open clusters, as \cite{Bonatto2006A&A...446..121B} ($14.8 \pm 2.4$ pc) and \cite{Joshi2007MNRAS.378..768J} ($17 \pm 3$\,pc) and more recently by \cite{CantatGaudin2020} who obtained ($15.3\pm 5.2$\,pc), who used stellar membership and distances obtained with data from the Gaia DR2 catalog for open clusters within 4\,kpc from the Sun.  It also agrees with results obtained with other classes of objects such as Cepheids \cite{Skowron2019} (see their Table S2) but is different from those obtained by other methods like using these high-latitude star counts ($27\pm4$\,pc) \cite{Chen2001}. We remark that this comparison should be taken with caution due to the influenced in the $R_{GC}$ by a combination of intrinsic scatter and the effects of warping, which we do not take into account.

We can also look at the distribution of cluster parameters as a function of height relative to the plane of the Galaxy. To accomplish this we obtain averages of the parameters $\log(age)$, $A_V$ and [Fe/H] for the clusters in bins determined from the distribution of heights to the plane. The following plots in Figure \ref{fig:Z-distribution} show the resulting variations that, despite the large dispersion, confirm general trends expected for the distribution of properties in the Galaxy. For instance, since the dust is strongly concentrated in the galactic plane, the clusters situated in the plane (small Z modulus) have larger mean Av, since their line-of sightS cross larger distances in regions of high density Interstellar Medium. On the other hand, it is known that stars (and clusters) situated at larger distances from the plane are older. This is confirmed by the first panel of Figure \ref{fig:Z-distribution}, where we see that only clusters older than 1\,Gyr reach distances $\geq$ 0.4\,kpc from the plane. It is usually accepted that the clusters are born in the Galactic plane, and then are scattered during their life to larger $|z|$ by collisions with molecular clouds \cite{Gustafsson}. On the other hand, the metallicity enrichment of the Galactic disk is a slow continuous process; as a result, old stars have low metallicities; combined with the scattering process that we just discussed, low metallicities are found in high $|z|$ stars, as confirmed by the last panel of Figure \ref{fig:Z-distribution}. This trend was also seen by \cite{FehZhightgrad} using data from the Apache Point Observatory Galactic Evolution Experiment data release 13 (DR13 hereafter) and Gaia Tycho-Gaia Astrometric Solution. Still another process that may contribute to having older clusters at high $|z|$ is that, although they oscillate and cross the Galactic plane, these clusters spend a larger fraction of their time in regions with smaller density of stars and of interstellar matter, which  guarantees a longer life.

\section{The spiral pattern rotation speed}

The method used in this work to estimate the value of $\Omega_p$ is the same presented in \cite{Dias2019}. Basically, from the birthplace position, we rotate forward this point an angle $\Omega_pT$ around the Galactic Center, and obtain a point situated on the present-day position of the same arm. We are assuming that the birthplace of a cluster represents a point of a spiral arm, a time $T$ ago, since it has been known 
for a long time that young open clusters are tracers of the spiral structure, which 
is also confirmed by Figure \ref{fig:arms-age}.

We consider that the shape of the spiral arms is conserved, so that every point of the initial spiral arm has a corresponding point on the present-day arm obtained by a rotation angle $\Delta\theta$ = $\theta_f$ - $\theta_i$. 
In this way, clusters with  different ages produce independent measures of the rotation velocity, according to the equation \ref{kinematics}, where the unknown parameter is $\Omega_p$, 

\begin{equation}
\theta_f = \theta_i + \Omega_{p}\times T
\label{kinematics}
\end{equation}
where, in polar coordinates, the azimuth $\theta_f$ is the present-day position angle of the arm, $\theta_i$ the birthplace position angle of a cluster, $\Omega_p$ the rotation velocity of the arms, and $T$ is the age of the cluster. \\

\subsection{The birthplace of the open clusters}

We find the birthplace of each cluster by integrating backward over their orbits for a time interval $T$ equal to their age, starting from the present-day initial conditions (positions and space velocities).
The numerical integration of the equations of motion was performed by means of a fifth-order Runge–Kutta integration procedure, with a typical time step of 0.1 Myr.

The birthplaces depend on the adopted rotation curve, which we present and discuss in the next section.
All the rotation curves have to be normalized to the adopted rotation velocity of the local standard of rest (LSR; $V_0$) and to the solar radius ($R_0$), of which the values $V_0$ = 240 km\,s$^{-1}$ and $R_0$ = 8.3 kpc were taken from \cite{Reid2016} and \cite{Gillessen2017}, respectively. 
To compute the heliocentric $U$ and $V$ velocities and the respective errors of each open cluster, we use the equatorial coordinates, distances, proper motions, and radial line-of-sight velocities, following the formalism described by \cite{JohnsonSoderblom1987}, and in order to pass to the LSR reference frame, we add the components of the solar motion $U_{\odot} = (-11.10\pm0.75)$\,km s$^{-1}$ and $V_{\odot} = (12.24\pm0.47)$\,km s$^{-1}$ (\cite{Schonrich2010}). 
The errors of $U_{LSR}$,$V_{LSR}$ velocities were determined by the usual propagation formula.

The errors in the  birthplace positions were estimated 
through Monte Carlo method, by integrating the orbits 1000 times from the 
generated Gaussian distributions, considering the values and uncertainties of distance, proper motion and velocity.  The birthplaces of the clusters
were taken as the mean of the distributions of the results of the
integrations, with the uncertainties represented by the standard
deviations. We obtain a typical error of about 140 pc in the birthplace positions.\\

\subsection{The Galactic rotation curve}

\begin{figure}
\centering
\includegraphics[scale = 1]{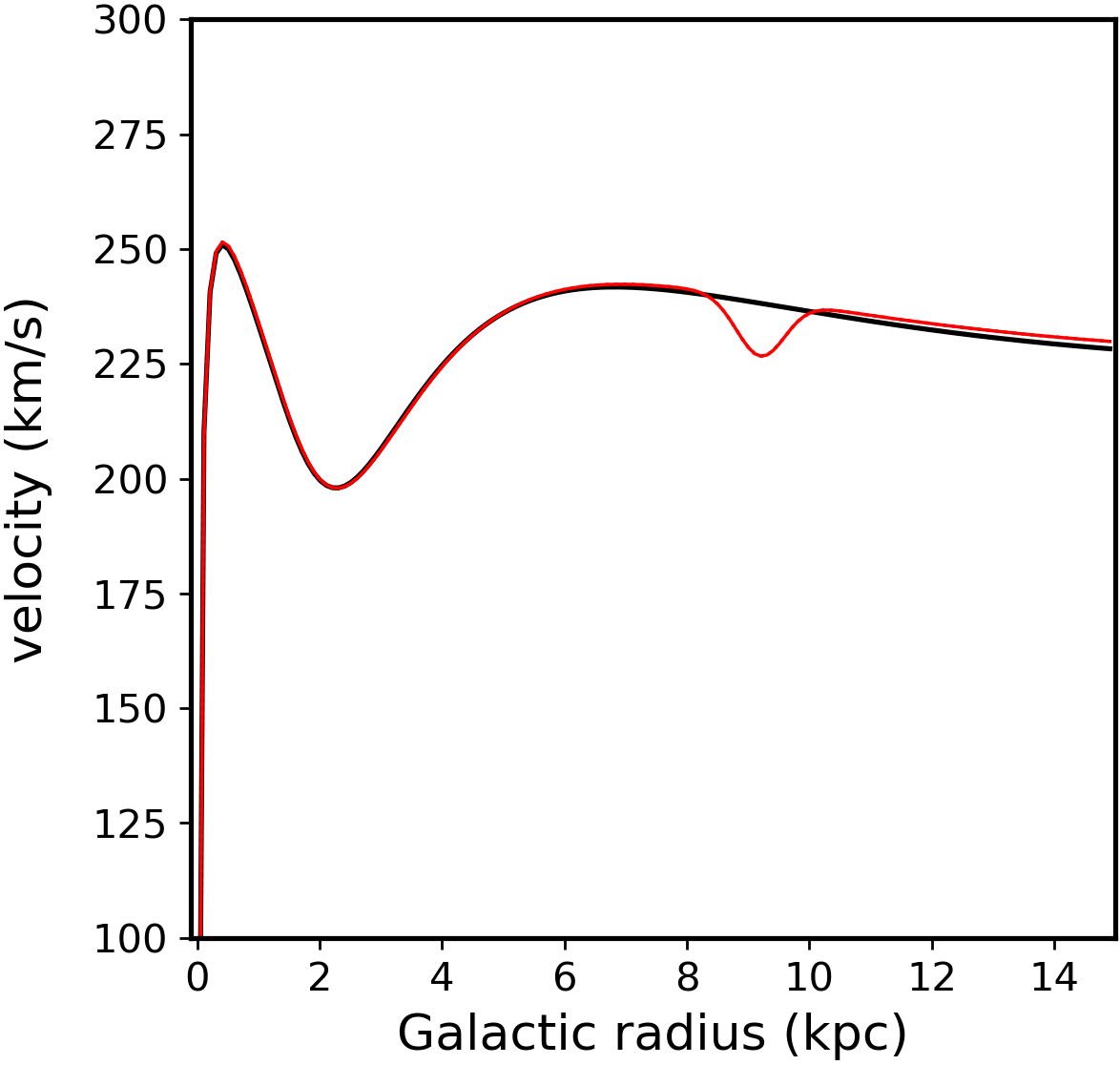}
\caption{Rotation curves of the Galaxy that can be classified as plane and plane with a dip centred at 9.2\,kpc. The analytical expressions for both rotation curves are given by Equations (2) and (3), respectively.}
\label{fig:rotation-curve}
\end{figure}

In this study, we used two analytical expressions for the rotation curve of the Galaxy, that are justified in terms of gravitational potential models of the Milky Way, and were fitted to different types of tracers, such as HI and CO tangent line data (\cite{Burton_Gordon1978,Clemens1985,Fich_Blitz_Stark1989}), and data of maser sources from VLBI observations (\cite{Reid2019,Rastorguev_etal2017}). One of the rotation curves is flat in the region of interest for the present work, the other one is not much different, but presents a narrow Gaussian dip (width\,0.39 kpc, depth 12.5\,km\,s$^{-1}$) centred at 9.2\,kpc.
In Figure \ref{fig:rotation-curve}, we present the rotation curves used in the present work. The flat rotation curve and the rotation curve with the dip at 9.2\,kpc are given by the following expressions (2) and (3), respectively:
\begin{eqnarray}
V_{\mathrm{rot}}(R)&=&298.9\,\mathrm{e}^{-\left(\frac{R}{4.73}\right)-\left(\frac{0.036}{R}\right)} + 231.8\,\mathrm{e}^{-\left(\frac{R}{1416.4}\right)-\left(\frac{3.72}{R}\right)^{2}} \\
V_{\mathrm{rot}}(R)&=&300.5\,\mathrm{e}^{-\left(\frac{R}{4.71}\right)-\left(\frac{0.037}{R}\right)} + 233.6\,\mathrm{e}^{-\left(\frac{R}{1539.3}\right)-\left(\frac{3.74}{R}\right)^{2}}-12.5\,\mathrm{e}^{-\frac{1}{2}\left(\frac{R-9.2}{0.39}\right)^{2}}\,,
\label{eq:Vrot}
\end{eqnarray}
with $R$ given in kpc and $V_{\mathrm{rot}}$ given in km\,s$^{-1}$. From the above rotation curves, we derive the radial gradient of the axysimmetric Galactic gravitational potential, $\partial\Phi_0/\partial R = V^2_{\rm rot}(R)/R$, used to integrate the orbits of the clusters and to determine their birthplaces.

For our sample,  the two curves did not yield  very different results. Note that the dip in the second rotation curve can be fitted adequately by different tracers (masers sources and HII regions) and can be explained as follows.  
The minimum in the rotation curve has contributions from both both  the density of  gas and of stars. 
In the case of gas, it is the "vacuum pump" effect promoted by the spiral potential. With little gas at corotation, then few stars will form there (except the portion of matter which must be stuck around the stable Lagrangian point and in the banana orbits). So that is why we see a minimum of density in young clusters. The older clusters, in principle, could fill the void of the corotation. However, we must introduce here the question of the secular transfer of angular momentum between stars and arms: stars within corotation give angular momentum to the arms, reducing their orbital radius and by the other hand, stars out of corotation gain angular momentum by the arms, which increases their orbiting radius. So the net effect on the stars is also a minimum of density in the corotation. This minimum is also seen with older clusters. The minimum of the rotation curve must really be offset from the minimum of density. The reader will find a detailed discussion of the issue in \cite{Barros2013}.
\\

\subsection{The zero-age arms}

Initially, we intended to use the log periodic spiral functions of \cite{Reid2014} and \cite{Reid2019} to trace zero-age arms positions. However, the aim of this work is not to
obtain a satisfactory fit over a large region of the Galaxy, but the best fit possible over the solar neighbourhood. In this way, we preferred to use polynomial fit to the position of the masers from \cite{Reid2014} and \cite{Reid2019}
which gives more liberty to the adjusted curve to run along the deviations from spirals. The work of \cite{Xu2006} shows strong deviations of the Carina arm from a spiral, and also indicates that the Perseus arm deviates from a pure logarithmic spiral, showing variable pitch angle. Note that the polynomial and logarithmic spirals fits are not in disagreement in the region of interest of our work as shown by \cite{Dias2019}.

We applied the same procedure to the four arms to be homogeneous, which gives a second-order polynomial for the Local and Sagittarius-Carina arms and sixth-order polynomial for the Perseus and Scutum arms. Clearly, in our method the values of rotation velocity of the spiral arms ($\Omega_p$) and the corotation radius depend on the location of the zero-age arms. Slightly different results can be obtained depending on the choice of the function representing the arms. The selection of the clusters belonging  to each arm was done considering the arms with width of $\pm$0.3 kpc to accommodate the different values estimated for each arm and the uncertainties associated with the present-day and birthplace position of the open clusters.

\section{Results and discussion}
\label{sec:results_discussion}

\begin{figure}
\centering
\includegraphics[scale = 0.8]{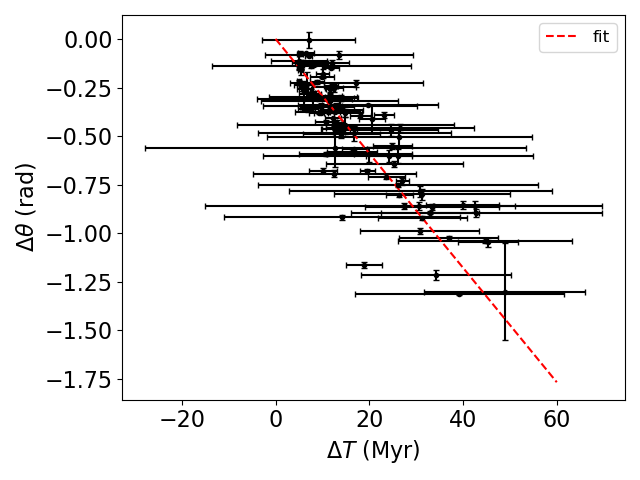}
\caption{ The angle between the birthplaces of the clusters, and the present day position of the corresponding spiral arm (zero age) at the same Galactic radius. The birthplaces 
were obtained using for integration the rotation curve without dip. The points represent the total sample. The slope of the linear fit to $\frac{\Delta\theta}{\Delta T}$ gives $\Omega_p = 28.5\pm0.8$ km\,s$^{-1}$\,kpc$^{-1}$, that we adopt as our final result.}

\label{fig:linearfit}
\end{figure}

Using the equation \ref{kinematics} we obtain the $\Omega_p$ by weighted linear least-squares fit, considering the uncertainty associated with each measurement as the weight. We applied the chi-square error statistic to determine the best angular coefficient to $\frac{\Delta\theta}{\Delta T}$, as presented in Figure \ref{fig:linearfit}. 

In Table \ref{tab:omegap}, we present the results of $\Omega_p$ obtained using both rotation curves, flat and with dip. The table also shows the $\Omega_p$ estimated for each arm. 
We found an excellent agreement with  the value of both $\Omega_p$ found in our previous work using the same method \cite{Dias2019}.
However, the values of $\Omega_p$ are not directly comparable with those published in \cite{Dias2005} since we used the rotation curves scaled to different values of $R_{0}$ and $V_{0}$.

The values of $\Omega_p$ from each rotation curve, presented in Table \ref{tab:omegap},  agree within the estimated errors, showing no statistical distinction. It can also be noted that the value obtained for each arm agree within the errors, which provides an observational evidence in favor of the classical stationary density wave theory where the arms rotates as a rigid body with the same velocity $\Omega_p$. The values are also in agreement with the range previously determined in the literature which range from ($20.1 \pm 5.3$) km\,s$^{-1}$\,kpc$^{-1}$ to ($25.2 \pm 0.5$) km\,s$^{-1}$\,kpc$^{-1}$ presented by \cite{Koda16} and \cite{Dambis15} respectively, although none use Gaia data.

The knowledge of $\Omega_p$ allows us to derive the corotation radius ($R_c$) by the relation $R_c = \frac{V_{rot}(R_c)}{\Omega_p}$. It is the radius in the Galactic disk where the stars rotate with the same velocity of the arms, that is, the strongest Galactic resonance.The value obtained in this work of ($28.5 \pm 1.0$) km\,s$^{-1}$\,kpc$^{-1}$ implies ($R_c = 8.42 \pm 0.46$) kpc, close to the solar Galactic orbit ($R_c/R_0 = 1.01\pm0.08$). 
It is in excellent agreement with those published in \cite{Dias2019} and \cite{Dias2005}. It is interesting to note that the corotation radius near the radius of the Local arm was not unexpected since, in the present-day positions (Figure \ref{fig:arms-age}) the young clusters lead the Sagittarius-Carina arm, are centred in the Local arm, and lag the Perseus arm.\\

\begin{table}
  \caption[]{$\Omega_p$ (in km\,s$^{-1}$\,kpc$^{-1}$) obtained with a plane Galactic rotation curve and with the same curve with a dip centered at 9.2 kpc. In this study we adopted c kpc and $V_{0}=240$ km\,s$^{-1}$. The method used is the linear fit to $\frac{\Delta\theta}{\Delta T}$.}   
\label{tab:omegap}
\begin{center}
\begin{tabular}{lcc}
\hline 
arm      & plane curve            &    curve with a dip \\ 
all      & $28.5\pm0.8$           &    $28.8\pm0.8$         \\     
Perseus  & $26.3\pm1.0$           &    $26.4\pm1.0$         \\     
Local    & $29.9\pm1.0$           &    $29.9\pm1.0$         \\     
Carina   & $29.8\pm2.7$           &    $29.8\pm2.7$         \\     
Scutum   & $23.7\pm1.0$           &    $26.1\pm0.8$         \\     
\hline
\end{tabular}
\end{center}
\end{table}

\subsection{The nature of the 9.5 kpc gap}
\begin{figure}[ht]
\centering
\includegraphics[scale = 0.6]{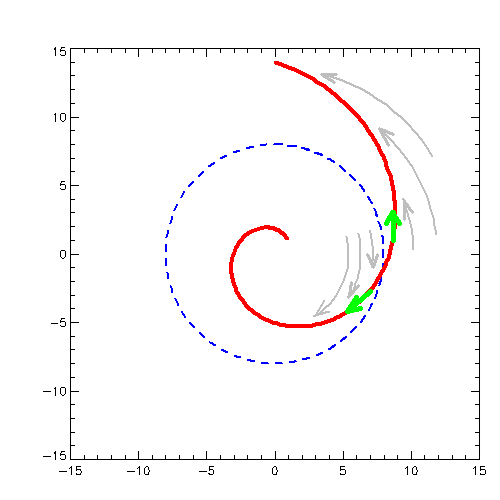}
\caption{A schematic model for the Galactic gas gap. A spiral arm is represented in red, it is a groove in the gravitational potential of the disk, which acts like a river for the interstellar gas.  Due to the rigid body-like rotation of the spiral arm, and to the flat rotation curve of the Galaxy, the interstellar gas which is being swept, represented by grey arrows, penetrates  the arms from opposite sides in the inner and outer regions of the disk, the frontier being the corotation radius, represented by the dashed blue circle. The angle of penetration of the gas determines the direction of the gas flow inside the arm. This results in opposite flow direction indicated by green arrows on the two sides of the corotation circle, a phenomenon which decreases the gas density in the arms  near the corotation radius.}
\centering
\label{fig:model-gap}
\end{figure}

The corotation divides the Galaxy into two parts, external and internal with different characteristics due to evolution over time, and at corotation  a minimum of density like a gap is expected.
The existence of this gap in the distribution of open clusters was already  known (eg. \cite{Barros2013}, where the adopted $R_0$ was 7.5 kpc), and is also present in the distribution of Cepheids \cite{Barros2013}. Thanks to the better age determinations available now, we have the opportunity to re-discuss the nature of this minimum. In Figure \ref{fig:RGC-distribution}, we present the distribution of Galactic radial distances for open clusters with ages smaller than about 50 Myr (log(age)$<$ 7.7), in which the gap can be seen to be much wider (about 1 kpc) than in the entire sample, without age selection. Our interpretation is that in the Galactic radius range 8.5-9.5 kpc, star formation and cluster formation are inhibited by the existence of a deep minimum in the gas density. A simple model of the gap phenomenon is shown in Figure \ref{fig:model-gap}, which helps to understand some basic aspects of it. We call attention
to the fact that the interstellar gas flows following circular orbits expressed by the rotation curve, in the inter-arm regions, 
but there is a component of flow in the radial direction, inside the arms, as shown in the simple model. Note that there is no radial mixing of the gas between the two halves of the disk, which favors independent chemical evolution of the two sides. 
A more professional hydrodynamic model is presented by \cite{Villegas2015} and the gas rarefaction at a radius situated a little farther than the solar radius 
has been effectively observed by means of the analysis of HI spectral profiles (\cite{Amores2013}).\\  

The metallicity as a function of Galactic radius is another interesting aspect to analyse. Here we are focusing on the general distribution of metallicity among the clusters, and not the value of  the individual metallicity estimate of any cluster in particular. 
The metallicity coincides with [Fe/H], the logarithm of iron abundance relative to hydrogen, normalized to the solar value. Although the present sample contains many more clusters, the general aspect of the distribution does not differ much from that which was obtained by (\cite{Lepine2011}, see their Figure 3). The metallicity data used in that paper were derived from high resolution spectroscopic observations of giant stars belonging to the clusters, which in some specific cases are more reliable than the isochrone fitting.  

\begin{figure}[h!]
\centering
\includegraphics[width=\textwidth]{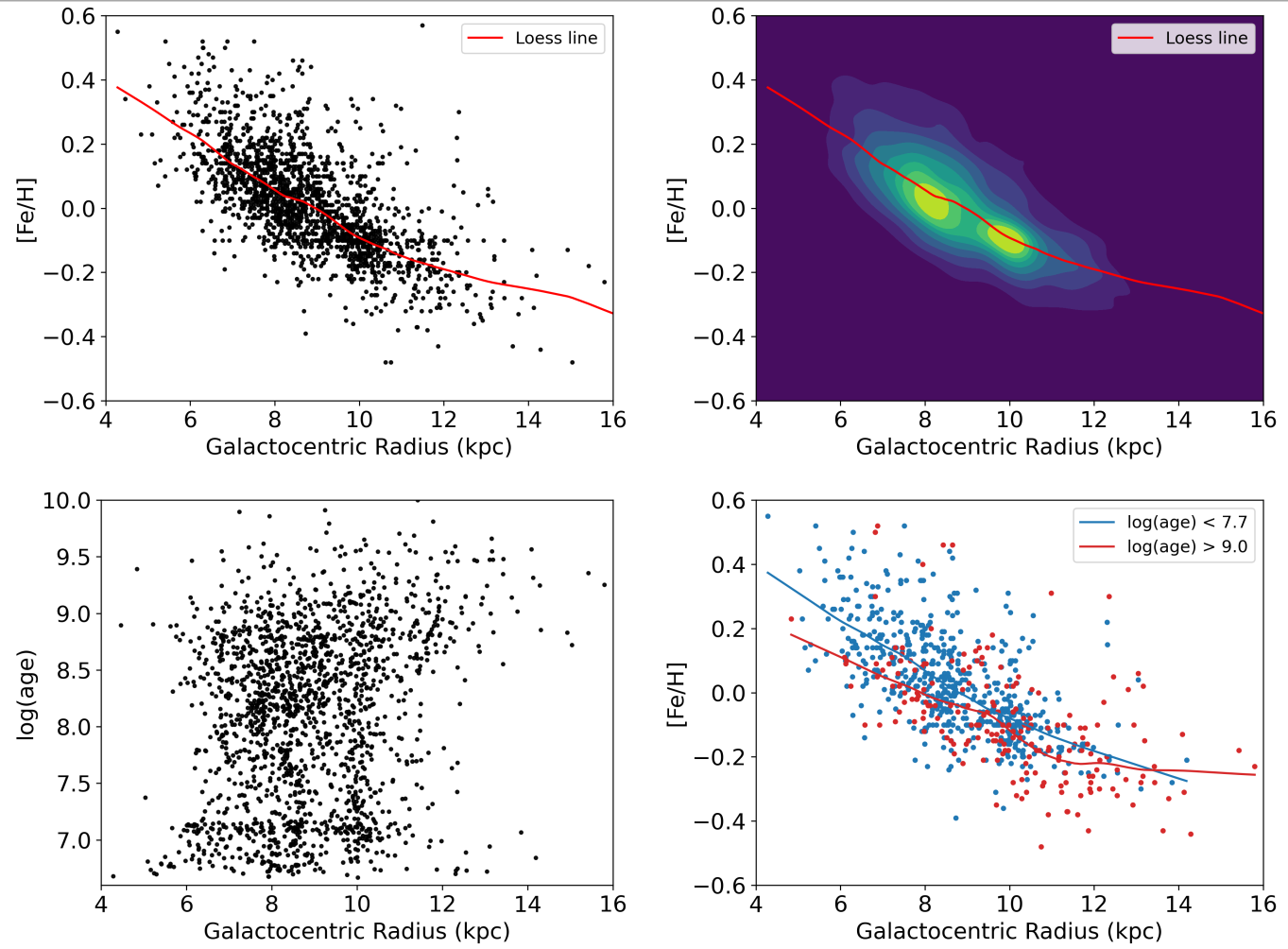}

\caption{(a) Metallicity, equal to iron abundance log [Fe/H] normalized to the Solar value as a function of Galactic radius. In the upper left panel we show all individual points with a non-parametric regression LOESS line over-plotted. In the upper right panel a kernel density estimate in ([Fe/H], radius) space is shown.  A gap at 9\,kpc can be seen, as well as a slight flattening of the gradient beyond 10\,kpc. In the lower left panel the logarithm of the age of the clusters is presented as a function of the Galactic radius, showing a prevalence of old clusters in the region with radius larger than 11\,kpc. In the lower right panel, the [Fe/H] distribution of the old and young population of clusters is shown with their respective non-parametric regression LOESS line over-plotted.}
\label{fig:metallicity}
\end{figure}

The main characteristics of the metallicity distribution are presented in Figure \ref{fig:metallicity}. In the upper left panel, we show all individual points with a non-parametric regression LOESS line over-plotted. In the upper right panel, a kernel density estimate in ([Fe/H], radius) space is shown.  A gap at 9 kpc can be seen, as well as a slight flattening of the gradient beyond 10 kpc. The flattening is not surprising since this is also a characteristic of the prior used for the metallicity as described in \cite{Monteiro2020}. In the lower left panel, the logarithm of the age of the clusters is presented as a function of the Galactic radius, showing a prevalence of old clusters at Galactic radii larger than 11 kpc. In the lower right panel, the distribution of the old and young population of clusters is shown with their respective non-parametric regression LOESS line over-plotted. The flattening of the gradient is more evident in the older population.

The large spread of radius at any given metallicity value is evident. The observed spread is in part due to the low precision of the metallicity estimates, but can also indicate a connection with stellar migrations and with the flow of gas inside spiral arms, understood as grooves in the Galactic potential. Note that at radius 8.5 kpc, for instance, we can find clusters with metallicities ranging from about -0.2 to +0.5! A detailed model is beyond the scope of this work, and large scale, precise data sets will be needed to draw any conclusions. Until the recent past, works on Galactic metallicity used to fit the metallicity distribution of open clusters or other tracers, by a single straight line going from about 6 to 12 kpc.  Although they give good overall results, these models do not help to understand the complexity of the transition zone related to corotation, which is now becoming more evident in the newer and larger data sets that are starting to be available. 

\begin{figure}
\centering
\includegraphics[scale = 0.55]{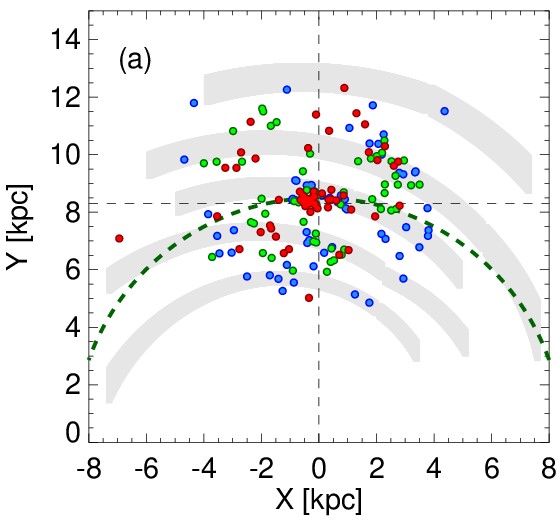}
\includegraphics[scale = 0.55]{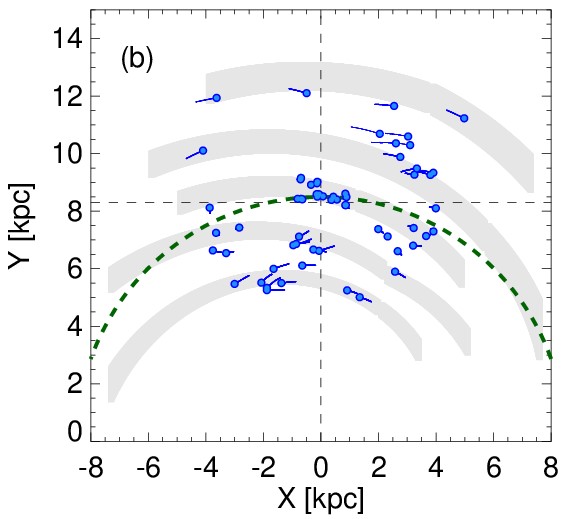}\\
\includegraphics[scale = 0.55]{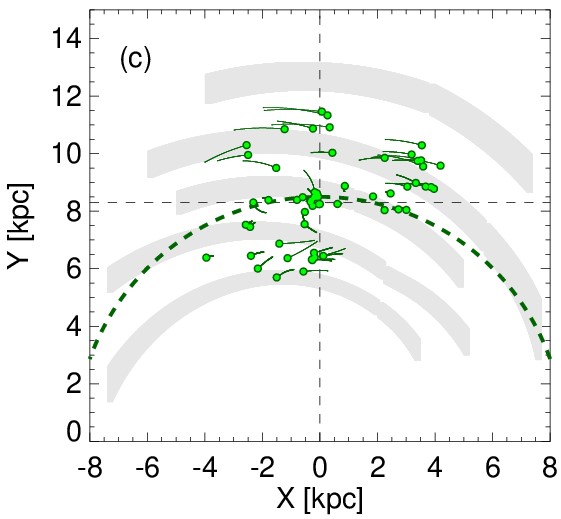}
\includegraphics[scale = 0.55]{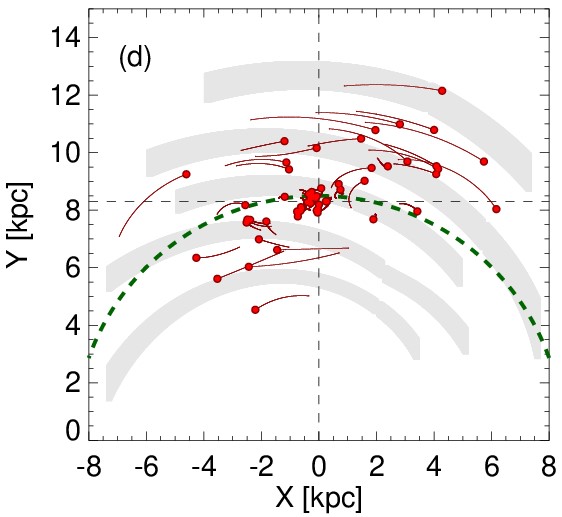}
\caption{Results of the backward integration of the orbits of open clusters younger than 50\,Myr (log(age)\,$<$\,7.7). The open clusters are separated in colors by the corresponding intervals of ages: blue for clusters with age\,$<$\,11.5\,Myr; green for clusters with 11.5\,$<$\,age\,$<$\,23.5\,Myr; and red for clusters with age\,$>$\,23.5\,Myr. The green dashed circle represents the corotation circle. The grey shaded areas show segments of the spiral arms from \cite{Reid2019}, which are, from top to down, respectively, the Outer arm, the Perseus arm, the Local arm, the Sagittarius-Carina arm, and the Scutum-Centaurus arm. The integration of the orbits was done in the reference frame of the spiral pattern with angular velocity \mbox{$\Omega_p = 28.5$\,km\,s$^{-1}$\,kpc$^{-1}$}. Panel (a) shows the present-day positions of the open clusters on the Galactic plane. Panel (b) shows the birthplaces of the open clusters younger than 11.5\,Myr (filled circles), with the tracing of the orbits since their present-day positions. Panels (c) and (d) are the same of panel (b) but for the birthplaces and orbits of the open clusters in the age intervals of 11.5\,$<$ age\,$<$\,23.5\,Myr and age\,$>$\,23.5\,Myr, respectively.}
\label{fig:orbits}
\end{figure}

In Figure \ref{fig:orbits}, we present the results of the backward integration of the orbits of open clusters younger than 50\,Myr (log(age)\,$<$\,7.7), with open clusters separated in colors corresponding to intervals of ages with sub-samples with approximately equal numbers of objects ($\sim$ 60 objects in each sub-sample): blue for clusters with age\,$<$\,11.5\,Myr; green for clusters with 11.5\,$<$\,age\,$<$\,23.5\,Myr; and red for clusters with age\,$>$\,23.5\,Myr.
Panel (a) shows the present-day positions of the clusters; panel (b) shows the birthplaces of the younger cluster sub-sample, along with the tracing of their orbits; panels (c) and (d) are the same of panel (b) but for the intermediate-age and older cluster sub-samples, respectively.  
Since the integration of the orbits was done in the reference frame of the spiral pattern, i.e., rotating with the angular velocity \mbox{$\Omega_p = 28.5$\,km\,s$^{-1}$\,kpc$^{-1}$}, following the sequence of panels, one can see the displacement of the clusters in opposite directions from the corotation circle (green dashed circle in the figure), especially in the older clusters subsample (clusters in red): inside the corotation circle, the clusters reach their birthplaces moving in the counterclockwise direction, approaching the segments of the Sagittarius-Carina and Scutum-Centaurus arms in the fourth quadrant of the Galaxy; outside the corotation circle, the clusters reach their birthplaces moving in the clockwise direction, approaching the Perseus and Outer arms segments in the second quadrant. The clusters near the corotation circle (located in the Local arm) show little displacements mainly due to their low relative angular motion with respect to the spiral arms. 

In the dynamical context, we understand that the spiral arms are like grooves in the gravitational potential of the Galactic disk, and they behave like channels along which there is a gas flow. The arms rotate like rigid bodies and the open clusters are born in the spiral arms, as discussed in this text. Most of the stars and of the molecular clouds in the inter-arm regions move in orbits that do not depart much from circular orbits around the Galactic Center.
Note that these objects with differential rotation in the disk define the rotation curve.
The rotation curve of the Galaxy crosses the rotation curve of the rigid spiral arm structure at a Galactic radius that we call the corotation radius, which we determined to be close to the radius of Solar orbit. 
A strong resonance develops at the corotation radius, which is able to capture stars and open clusters in local libration orbits. As the spiral arms sweeps the interstellar medium due to its rotation, the interaction of the interstellar gas with arm's grooves induces gas flows in the arm. The gas flow runs along the arms, in opposite directions on each side (inner or outer) of the corotation radius (see Figure \ref{fig:model-gap}). This is at the origin of a low gas density or a gap, in the interior of the arms, at the corotation radius. This gap due the corotation and the gas flow in the arms are able to explain some aspects of the metallicity distribution in the Galactic plane.\\

\section{Conclusions}

We have presented the main characteristics of the recently available sample of 1750 open clusters, with added radial velocities and high quality of isochrone fitting and age determination. Based on a sample of dozens young clusters, we determined the rotation velocity of the spiral structure, assuming that the clusters are born 
in the spiral arms of the Galaxy. We recover the birthplaces by integrating backwards the orbits starting from the present position of the cluster, for a time equal to the age of the cluster.  Therefore, each birthplace represents the position of the arm at a given epoch, and we compare it with the zero-age arm position of the present day spiral arm traced by a sample of masers. 
The method applied consists in getting the rotation angle of the spiral 
arm needed to make it coincide with the present day structure. Each cluster sets out one rotation angle and one time interval (the age).
The set of birthplaces and actual positions of the same arms  gives the rate of change of angular position of the arms as a function of time. This allows  determination of the spiral pattern rotation speed of the Galaxy as  $\Omega_p = 28.5\pm1.0$ km\,s$^{-1}$\,kpc$^{-1}$, now including the Scutum arm in the analysis.

Interestingly, our analysis points out that the values of $\Omega_p$ estimated for each arm are similar within 1 $\sigma$ level error.
This  means that the arms do not have any  significant angular velocity  with respect to the others. It is an observational evidence that goes in the same direction of the first historical works on density wave theory (\cite{Lindblad1938} and \cite{LinShu1964}) in which the spiral pattern rotates as a rigid body with angular velocity $\Omega_p$. However, these historical works are not used any more  in present day research. The interpretation of the nature of the spiral arms that we adopt, in terms of grooves in the galactic gravitational potential of the disk \cite{Junqueira2013}, is quite different from the classical theory, but shares with it the rigid rotation of the spiral structure.

As expected from visual inspection of the positions of the young clusters in relation to  the zero-age position of the spiral arms, the corotation radius location is very close to the Solar orbit radius. In this work, the corotation radius, derived  from the value of $\Omega_p$, is ($R_c = 8.57\pm0.07$)\,kpc, while the ratio between the corotation radius and the solar radius is ($R_c/R_0 = 1.01\pm0.08$). 
We show evidence for a gap in the gas distribution of the Galactic disk at at about $9\,kpc$. The corotation radius result  supports the explanation that we give for the  gap  and for the strong change of metallicity gradient in the Galactic disk in the vicinity of the Sun.  The metallicity gradient is highlighted in Figure \ref{fig:metallicity}, a kernel density plot of distribution of [Fe/H] as a function of the Galactic radius.

The results are based on the unprecedented quality of the Gaia DR2 data, and on direct dynamical and chemical analyses that do not involve any complicated or polemical theory. We plan to update the present analyses based on Gaia EDR3, and on the resulting  updates of the cluster catalogues. In our opinion, this work contributes to understand the extended solar neighborhood in the context of Galactic dynamics, and will not suffer conceptual changes due to Gaia DR3.

\section*{Acknowledgments}
This research was performed using the facilities of the Laborat\'orio de Astrof\'isica Computacional da Universidade Federal de Itajub\'a (LAC-UNIFEI)
This work has made use of data from the European Space Agency (ESA) mission {\it Gaia} (\url{https://www.cosmos.esa.int/gaia}), processed by the {\it Gaia} Data Processing and Analysis Consortium (DPAC, \url{https://www.cosmos.esa.int/web/gaia/dpac/consortium}).
H. Monteiro would like to thank FAPEMIG grants APQ-02030-10 and CEX-PPM-00235-12. W.S.Dias acknowledges the S\~ao Paulo State Agency
FAPESP (fellowship 2013/01115-6). JRDL was supported by the Brazilian agency Conselho Nacional de Desenvolvimento Cient\'ifico e Tecnol\'ogico (CNPq)  grant 302546/2004-9.
We thank the referee for many comments that improved the text.

\bibliographystyle{frontiersinHLTH_FPHY}
\bibliography{refs2} 

\begin{thebibliography}{46}
\expandafter\ifx\csname natexlab\endcsname\relax\def\natexlab#1{#1}\fi
\expandafter\ifx\csname urlstyle\endcsname\relax
  \expandafter\ifx\csname doi\endcsname\relax
  \def\doi#1{doi:\discretionary{}{}{}#1}\fi \else
  \expandafter\ifx\csname doi\endcsname\relax
  \def\doi{doi:\discretionary{}{}{}\begingroup \urlstyle{rm}\Url}\fi \fi
\expandafter\ifx\csname selectlanguage\endcsname\relax
  \def\selectlanguage#1{}\fi

\bibitem[{{Gaia Collaboration} et~al.(2018){Gaia Collaboration}, {Brown},
  {Vallenari}, {Prusti}, {de Bruijne}, {Babusiaux} et~al.}]{GaiaCollab2018}
{Gaia Collaboration}, {Brown} AGA, {Vallenari} A, {Prusti} T, {de Bruijne} JHJ,
  {Babusiaux} C, et~al.
\newblock {Gaia Data Release 2. Summary of the contents and survey properties}.
\newblock {\em A\&A\/} {\bf 616} (2018) A1.
\newblock \doi{10.1051/0004-6361/201833051}.

\bibitem[{{Cantat-Gaudin} et~al.(2020){Cantat-Gaudin}, {Anders},
  {Castro-Ginard}, {Jordi}, {Romero-G{\'o}mez}, {Soubiran}
  et~al.}]{CantatGaudin2020}
{Cantat-Gaudin} T, {Anders} F, {Castro-Ginard} A, {Jordi} C, {Romero-G{\'o}mez}
  M, {Soubiran} C, et~al.
\newblock {Painting a portrait of the Galactic disc with its stellar clusters}.
\newblock {\em A\&A\/} {\bf 640} (2020) A1.
\newblock \doi{10.1051/0004-6361/202038192}.

\bibitem[{{Castro-Ginard} et~al.(2020){Castro-Ginard}, {Jordi}, {Luri},
  {{\'A}lvarez Cid-Fuentes}, {Casamiquela}, {Anders} et~al.}]{CastroGinard2020}
{Castro-Ginard} A, {Jordi} C, {Luri} X, {{\'A}lvarez Cid-Fuentes} J,
  {Casamiquela} L, {Anders} F, et~al.
\newblock {Hunting for open clusters in Gaia DR2: 582 new open clusters in the
  Galactic disc}.
\newblock {\em A\&A\/} {\bf 635} (2020) A45.
\newblock \doi{10.1051/0004-6361/201937386}.

\bibitem[{{Castro-Ginard} et~al.(2019){Castro-Ginard}, {Jordi}, {Luri},
  {Cantat-Gaudin}, and {Balaguer-N{\'u}{\~n}ez}}]{CastroGinard2019}
{Castro-Ginard} A, {Jordi} C, {Luri} X, {Cantat-Gaudin} T,
  {Balaguer-N{\'u}{\~n}ez} L.
\newblock {Hunting for open clusters in Gaia DR2: the Galactic anticentre}.
\newblock {\em A\&A\/} {\bf 627} (2019) A35.
\newblock \doi{10.1051/0004-6361/201935531}.

\bibitem[{{Liu} and {Pang}(2019)}]{LiuPang2019}
{Liu} L, {Pang} X.
\newblock {A Catalog of Newly Identified Star Clusters in Gaia DR2}.
\newblock {\em ApJS\/} {\bf 245} (2019) 32.
\newblock \doi{10.3847/1538-4365/ab530a}.

\bibitem[{{Sim} et~al.(2019){Sim}, {Lee}, {Ann}, and {Kim}}]{Sim2019}
{Sim} G, {Lee} SH, {Ann} HB, {Kim} S.
\newblock {207 New Open Star Clusters within 1 kpc from Gaia Data Release 2}.
\newblock {\em Journal of Korean Astronomical Society\/} {\bf 52} (2019)
  145--158.
\newblock \doi{10.5303/JKAS.2019.52.5.145}.

\bibitem[{{Ferreira} et~al.(2020){Ferreira}, {Corradi}, {Maia}, {Angelo}, and
  {Santos}}]{Ferreira2020}
{Ferreira} FA, {Corradi} WJB, {Maia} FFS, {Angelo} MS, {Santos} J J~F~C.
\newblock {Discovery and astrophysical properties of Galactic open clusters in
  dense stellar fields using Gaia DR2}.
\newblock {\em MNRAS\/} {\bf 496} (2020) 2021--2038.
\newblock \doi{10.1093/mnras/staa1684}.

\bibitem[{{Monteiro} et~al.(2020){Monteiro}, {Dias}, {Moitinho},
  {Cantat-Gaudin}, {L{\'e}pine}, {Carraro} et~al.}]{Monteiro2020}
{Monteiro} H, {Dias} WS, {Moitinho} A, {Cantat-Gaudin} T, {L{\'e}pine} JRD,
  {Carraro} G, et~al.
\newblock {Fundamental parameters for 45 open clusters with Gaia DR2, an
  improved extinction correction and a metallicity gradient prior}.
\newblock {\em MNRAS\/} {\bf 499} (2020) 1874--1889.
\newblock \doi{10.1093/mnras/staa2983}.

\bibitem[{{Junqueira} et~al.(2013){Junqueira}, {L{\'e}pine}, {Braga}, and
  {Barros}}]{Junqueira2013}
{Junqueira} TC, {L{\'e}pine} JRD, {Braga} CAS, {Barros} DA.
\newblock {A new model for gravitational potential perturbations in disks of
  spiral galaxies. An application to our Galaxy}.
\newblock {\em A\&A\/} {\bf 550} (2013) A91.
\newblock \doi{10.1051/0004-6361/201219769}.

\bibitem[{{Dias} et~al.(2002){Dias}, {Alessi}, {Moitinho}, and
  {L{\'e}pine}}]{Dias2002}
{Dias} WS, {Alessi} BS, {Moitinho} A, {L{\'e}pine} JRD.
\newblock {New catalogue of optically visible open clusters and candidates}.
\newblock {\em A\&A\/} {\bf 389} (2002) 871--873.
\newblock \doi{10.1051/0004-6361:20020668}.

\bibitem[{{Dias} et~al.(2021){Dias}, {Monteiro}, {L{\'e}pine}, and
  {Barros}}]{Dias2021}
{Dias} WS, {Monteiro} H, {L{\'e}pine} JRD, {Barros} DA.
\newblock {Updated parameters of 1758 open clusters based on Gaia DR2}.
\newblock {\em MNRAS\/} {\bf 486} (2021) 5726--5736.

\bibitem[{{Cantat-Gaudin} et~al.(2018){Cantat-Gaudin}, {Jordi}, {Vallenari},
  {Bragaglia}, {Balaguer-N{\'u}{\~n}ez}, {Soubiran} et~al.}]{cantat2018}
{Cantat-Gaudin} T, {Jordi} C, {Vallenari} A, {Bragaglia} A,
  {Balaguer-N{\'u}{\~n}ez} L, {Soubiran} C, et~al.
\newblock {A Gaia DR2 view of the open cluster population in the Milky Way}.
\newblock {\em A\&A\/} {\bf 618} (2018) A93.
\newblock \doi{10.1051/0004-6361/201833476}.

\bibitem[{{Krone-Martins} and
  {Moitinho}(2014)}]{Krone-Martins2014A&A...561A..57K}
{Krone-Martins} A, {Moitinho} A.
\newblock {UPMASK: unsupervised photometric membership assignment in stellar
  clusters}.
\newblock {\em A\&A\/} {\bf 561} (2014) A57.
\newblock \doi{10.1051/0004-6361/201321143}.

\bibitem[{{Dias} et~al.(2014){Dias}, {Monteiro}, {Caetano}, {L{\'e}pine},
  {Assafin}, and {Oliveira}}]{Dias2014}
{Dias} WS, {Monteiro} H, {Caetano} TC, {L{\'e}pine} JRD, {Assafin} M,
  {Oliveira} AF.
\newblock {Proper motions of the optically visible open clusters based on the
  UCAC4 catalog}.
\newblock {\em A\&A\/} {\bf 564} (2014) A79.
\newblock \doi{10.1051/0004-6361/201323226}.

\bibitem[{{Barford}(1985)}]{Barford1985}
{Barford} NC.
\newblock {\em {Experimental measurements: Precision, error and truth}\/}
  (Chichester: Wiley, 1985, 2nd ed.) (1985).

\bibitem[{{Carrera} et~al.(2019){Carrera}, {Bragaglia}, {Cantat-Gaudin},
  {Vallenari}, {Balaguer-N{\'u}{\~n}ez}, {Bossini} et~al.}]{Carrera2019}
{Carrera} R, {Bragaglia} A, {Cantat-Gaudin} T, {Vallenari} A,
  {Balaguer-N{\'u}{\~n}ez} L, {Bossini} D, et~al.
\newblock {Open clusters in APOGEE and GALAH. Combining Gaia and ground-based
  spectroscopic surveys}.
\newblock {\em A\&A\/} {\bf 623} (2019) A80.
\newblock \doi{10.1051/0004-6361/201834546}.

\bibitem[{{Soubiran} et~al.(2018){Soubiran}, {Cantat-Gaudin},
  {Romero-G{\'o}mez}, {Casamiquela}, {Jordi}, {Vallenari}
  et~al.}]{Soubiran2018}
{Soubiran} C, {Cantat-Gaudin} T, {Romero-G{\'o}mez} M, {Casamiquela} L, {Jordi}
  C, {Vallenari} A, et~al.
\newblock {Open cluster kinematics with Gaia DR2}.
\newblock {\em A\&A\/} {\bf 619} (2018) A155.
\newblock \doi{10.1051/0004-6361/201834020}.

\bibitem[{{Am{\^o}res} and {L{\'e}pine}(2005)}]{Amores2005}
{Am{\^o}res} EB, {L{\'e}pine} JRD.
\newblock {Models for Interstellar Extinction in the Galaxy}.
\newblock {\em AJ\/} {\bf 130} (2005) 659--673.
\newblock \doi{10.1086/430957}.

\bibitem[{{Reid} et~al.(2014){Reid}, {Menten}, {Brunthaler}, {Zheng}, {Dame},
  {Xu} et~al.}]{Reid2014}
{Reid} MJ, {Menten} KM, {Brunthaler} A, {Zheng} XW, {Dame} TM, {Xu} Y, et~al.
\newblock {Trigonometric Parallaxes of High Mass Star Forming Regions: The
  Structure and Kinematics of the Milky Way}.
\newblock {\em ApJ\/} {\bf 783} (2014) 130.
\newblock \doi{10.1088/0004-637X/783/2/130}.

\bibitem[{{Binney} and {Tremaine}(1987)}]{Binney}
{Binney} J, {Tremaine} S.
\newblock {\em {Galactic dynamics}\/} (Princeton, N.J. : Princeton University
  Press, c1987) (1987).

\bibitem[{{Am{\^o}res} et~al.(2013){Am{\^o}res}, {L{\'o}pez-Corredoira},
  {Gonz{\'a}lez-Fern{\'a}ndez}, {Moitinho}, {Minniti}, and
  {Gurovich}}]{Amores2013}
{Am{\^o}res} EB, {L{\'o}pez-Corredoira} M, {Gonz{\'a}lez-Fern{\'a}ndez} C,
  {Moitinho} A, {Minniti} D, {Gurovich} S.
\newblock {The long bar as seen by the VVV Survey. II. Star counts}.
\newblock {\em A\&A\/} {\bf 559} (2013) A11.
\newblock \doi{10.1051/0004-6361/201219846}.

\bibitem[{{Bonatto} et~al.(2006){Bonatto}, {Kerber}, {Bica}, and
  {Santiago}}]{Bonatto2006A&A...446..121B}
{Bonatto} C, {Kerber} LO, {Bica} E, {Santiago} BX.
\newblock {Probing disk properties with open clusters}.
\newblock {\em A\&A\/} {\bf 446} (2006) 121--135.
\newblock \doi{10.1051/0004-6361:20053573}.

\bibitem[{{Joshi}(2007)}]{Joshi2007MNRAS.378..768J}
{Joshi} YC.
\newblock {Displacement of the Sun from the Galactic plane}.
\newblock {\em MNRAS\/} {\bf 378} (2007) 768--776.
\newblock \doi{10.1111/j.1365-2966.2007.11831.x}.

\bibitem[{{Skowron} et~al.(2019){Skowron}, {Skowron}, {Mr{\'o}z}, {Udalski},
  {Pietrukowicz}, {Soszy{\'n}ski} et~al.}]{Skowron2019}
{Skowron} DM, {Skowron} J, {Mr{\'o}z} P, {Udalski} A, {Pietrukowicz} P,
  {Soszy{\'n}ski} I, et~al.
\newblock {A three-dimensional map of the Milky Way using classical Cepheid
  variable stars}.
\newblock {\em Science\/} {\bf 365} (2019) 478--482.
\newblock \doi{10.1126/science.aau3181}.

\bibitem[{{Chen} et~al.(2001){Chen}, {Stoughton}, {Smith}, {Uomoto}, {Pier},
  {Yanny} et~al.}]{Chen2001}
{Chen} B, {Stoughton} C, {Smith} JA, {Uomoto} A, {Pier} JR, {Yanny} B, et~al.
\newblock {Stellar Population Studies with the SDSS. I. The Vertical
  Distribution of Stars in the Milky Way}.
\newblock {\em ApJ\/} {\bf 553} (2001) 184--197.
\newblock \doi{10.1086/320647}.

\bibitem[{{Gustafsson} et~al.(2016){Gustafsson}, {Church}, {Davies}, and
  {Rickman}}]{Gustafsson}
{Gustafsson} B, {Church} RP, {Davies} MB, {Rickman} H.
\newblock {Gravitational scattering of stars and clusters and the heating of
  the Galactic disk}.
\newblock {\em A\&A\/} {\bf 593} (2016) A85.
\newblock \doi{10.1051/0004-6361/201423916}.

\bibitem[{{Li} et~al.(2018){Li}, {Zhao}, {Zhai}, and {Jia}}]{FehZhightgrad}
{Li} C, {Zhao} G, {Zhai} M, {Jia} Y.
\newblock {The Formation and Evolution of Galactic Disks with APOGEE and the
  Gaia Survey}.
\newblock {\em ApJ\/} {\bf 860} (2018) 53.
\newblock \doi{10.3847/1538-4357/aac50f}.

\bibitem[{{Dias} et~al.(2019){Dias}, {Monteiro}, {L{\'e}pine}, and
  {Barros}}]{Dias2019}
{Dias} WS, {Monteiro} H, {L{\'e}pine} JRD, {Barros} DA.
\newblock {The spiral pattern rotation speed of the Galaxy and the corotation
  radius with Gaia DR2}.
\newblock {\em MNRAS\/} {\bf 486} (2019) 5726--5736.
\newblock \doi{10.1093/mnras/stz1196}.

\bibitem[{{Reid} and {Dame}(2016)}]{Reid2016}
{Reid} MJ, {Dame} TM.
\newblock {On the Rotation Speed of the Milky Way Determined from H I
  Emission}.
\newblock {\em ApJ\/} {\bf 832} (2016) 159.
\newblock \doi{10.3847/0004-637X/832/2/159}.

\bibitem[{{Gillessen} et~al.(2017){Gillessen}, {Plewa}, {Eisenhauer}, {Sari},
  {Waisberg}, {Habibi} et~al.}]{Gillessen2017}
{Gillessen} S, {Plewa} PM, {Eisenhauer} F, {Sari} R, {Waisberg} I, {Habibi} M,
  et~al.
\newblock {An Update on Monitoring Stellar Orbits in the Galactic Center}.
\newblock {\em ApJ\/} {\bf 837} (2017) 30.
\newblock \doi{10.3847/1538-4357/aa5c41}.

\bibitem[{{Johnson} and {Soderblom}(1987)}]{JohnsonSoderblom1987}
{Johnson} DRH, {Soderblom} DR.
\newblock {Calculating Galactic Space Velocities and Their Uncertainties, with
  an Application to the Ursa Major Group}.
\newblock {\em AJ\/} {\bf 93} (1987) 864.
\newblock \doi{10.1086/114370}.

\bibitem[{{Sch{\"o}nrich} et~al.(2010){Sch{\"o}nrich}, {Binney}, and
  {Dehnen}}]{Schonrich2010}
{Sch{\"o}nrich} R, {Binney} J, {Dehnen} W.
\newblock {Local kinematics and the local standard of rest}.
\newblock {\em MNRAS\/} {\bf 403} (2010) 1829--1833.
\newblock \doi{10.1111/j.1365-2966.2010.16253.x}.

\bibitem[{{Burton} and {Gordon}(1978)}]{Burton_Gordon1978}
{Burton} WB, {Gordon} MA.
\newblock {Carbon monoxide in the Galaxy. III - The overall nature of its
  distribution in the equatorial plane}.
\newblock {\em A\&A\/} {\bf 63} (1978) 7--27.

\bibitem[{{Clemens}(1985)}]{Clemens1985}
{Clemens} DP.
\newblock {Massachusetts-Stony Brook Galactic plane CO survey - The Galactic
  disk rotation curve}.
\newblock {\em ApJ\/} {\bf 295} (1985) 422--428.
\newblock \doi{10.1086/163386}.

\bibitem[{{Fich} et~al.(1989){Fich}, {Blitz}, and
  {Stark}}]{Fich_Blitz_Stark1989}
{Fich} M, {Blitz} L, {Stark} AA.
\newblock {The rotation curve of the Milky Way to 2 R(0)}.
\newblock {\em ApJ\/} {\bf 342} (1989) 272--284.
\newblock \doi{10.1086/167591}.

\bibitem[{{Reid} et~al.(2019){Reid}, {Menten}, {Brunthaler}, {Zheng}, {Dame},
  {Xu} et~al.}]{Reid2019}
{Reid} MJ, {Menten} KM, {Brunthaler} A, {Zheng} XW, {Dame} TM, {Xu} Y, et~al.
\newblock {Trigonometric Parallaxes of High-mass Star-forming Regions: Our View
  of the Milky Way}.
\newblock {\em ApJ\/} {\bf 885} (2019) 131.
\newblock \doi{10.3847/1538-4357/ab4a11}.

\bibitem[{{Rastorguev} et~al.(2017){Rastorguev}, {Utkin}, {Zabolotskikh},
  {Dambis}, {Bajkova}, and {Bobylev}}]{Rastorguev_etal2017}
{Rastorguev} AS, {Utkin} ND, {Zabolotskikh} MV, {Dambis} AK, {Bajkova} AT,
  {Bobylev} VV.
\newblock {Galactic masers: Kinematics, spiral structure and the disk dynamic
  state}.
\newblock {\em Astrophysical Bulletin\/} {\bf 72} (2017) 122--140.
\newblock \doi{10.1134/S1990341317020043}.

\bibitem[{{Barros} et~al.(2013){Barros}, {L{\'e}pine}, and
  {Junqueira}}]{Barros2013}
{Barros} DA, {L{\'e}pine} JRD, {Junqueira} TC.
\newblock {A Galactic ring of minimum stellar density near the solar orbit
  radius}.
\newblock {\em MNRAS\/} {\bf 435} (2013) 2299--2321.
\newblock \doi{10.1093/mnras/stt1454}.

\bibitem[{{Xu} et~al.(2006){Xu}, {Reid}, {Zheng}, and {Menten}}]{Xu2006}
{Xu} Y, {Reid} MJ, {Zheng} XW, {Menten} KM.
\newblock {The Distance to the Perseus Spiral Arm in the Milky Way}.
\newblock {\em Science\/} {\bf 311} (2006) 54--57.
\newblock \doi{10.1126/science.1120914}.

\bibitem[{{Dias} and {L{\'e}pine}(2005)}]{Dias2005}
{Dias} WS, {L{\'e}pine} JRD.
\newblock {Direct Determination of the Spiral Pattern Rotation Speed of the
  Galaxy}.
\newblock {\em ApJ\/} {\bf 629} (2005) 825--831.
\newblock \doi{10.1086/431456}.

\bibitem[{{Koda} et~al.(2016){Koda}, {Scoville}, and {Heyer}}]{Koda16}
{Koda} J, {Scoville} N, {Heyer} M.
\newblock {Evolution of Molecular and Atomic Gas Phases in the Milky Way}.
\newblock {\em ApJ\/} {\bf 823} (2016) 76.
\newblock \doi{10.3847/0004-637X/823/2/76}.

\bibitem[{{Dambis} et~al.(2015){Dambis}, {Berdnikov}, {Efremov}, {Kniazev},
  {Rastorguev}, {Glushkova} et~al.}]{Dambis15}
{Dambis} AK, {Berdnikov} LN, {Efremov} YN, {Kniazev} AY, {Rastorguev} AS,
  {Glushkova} EV, et~al.
\newblock {Classical Cepheids and the spiral structure of the milky way}.
\newblock {\em Astronomy Letters\/} {\bf 41} (2015) 489--500.
\newblock \doi{10.1134/S1063773715090017}.

\bibitem[{{P{\'e}rez-Villegas} et~al.(2015){P{\'e}rez-Villegas}, {G{\'o}mez},
  and {Pichardo}}]{Villegas2015}
{P{\'e}rez-Villegas} A, {G{\'o}mez} GC, {Pichardo} B.
\newblock {The galactic branches as a possible evidence for transient spiral
  arms}.
\newblock {\em MNRAS\/} {\bf 451} (2015) 2922--2932.
\newblock \doi{10.1093/mnras/stv1157}.

\bibitem[{{L{\'e}pine} et~al.(2011){L{\'e}pine}, {Cruz}, {Scarano}, {Barros},
  {Dias}, {Pomp{\'e}ia} et~al.}]{Lepine2011}
{L{\'e}pine} JRD, {Cruz} P, {Scarano} J S, {Barros} DA, {Dias} WS,
  {Pomp{\'e}ia} L, et~al.
\newblock {Overlapping abundance gradients and azimuthal gradients related to
  the spiral structure of the Galaxy}.
\newblock {\em MNRAS\/} {\bf 417} (2011) 698--708.
\newblock \doi{10.1111/j.1365-2966.2011.19314.x}.

\bibitem[{{Lindblad}(1938)}]{Lindblad1938}
{Lindblad} B.
\newblock {On the theory of spiral structure in the nebulae.}
\newblock {\em ZAP\/} {\bf 15} (1938) 124.

\bibitem[{{Lin} and {Shu}(1964)}]{LinShu1964}
{Lin} CC, {Shu} FH.
\newblock {On the Spiral Structure of Disk Galaxies.}
\newblock {\em ApJ\/} {\bf 140} (1964) 646.
\newblock \doi{10.1086/147955}.

\end{thebibliography}

\end{document}